\documentclass[10pt]{article}

\RequirePackage{amsmath}[1997/03/20]
\RequirePackage{amssymb}
\RequirePackage{amsthm}
\RequirePackage{graphicx}
\RequirePackage{cite}
\RequirePackage{bm}
\RequirePackage{url}

\begin{document}

\twocolumn[
\begin{center}
{\LARGE
Second-level randomness test based on the Kolmogorov-Smirnov test
}

\bigskip

{\large 
Akihiro Yamaguchi$^{1}$ and Asaki Saito$^{2}$ \par
}
\bigskip

\end{center}

\hspace{1.5cm} $^1$ Fukuoka Institute of Technology, 3-30-1 Wajiro-higashi, Higashi-ku, Fukuoka 811-0295, Japan \\

\vspace{-0.4cm}

\hspace{1.5cm} $^2$ Future University Hakodate, 116-2 Kamedanakano-cho, Hakodate, Hokkaido 041-8655, Japan \\

\vspace{-0.4cm}

\begin{center}
\abstract
\end{center}
We analyzed the effect of the deviation of the exact distribution of the p-values from the uniform distribution on the Kolmogorov-Smirnov (K-S) test that was implemented as the second-level randomness test. 
We derived an inequality that provides an upper bound on the expected value of the K-S test statistic when the distribution of the null hypothesis differs from the  exact distribution.
Furthermore, we proposed a second-level test based on the two-sample K-S test with an ideal empirical distribution as a candidate for improvement.  
\\
\\
{\bf Keywords}: Kolmogorov-Smilrov test, uniformity of p-value,  chaotic true orbits
\\
\\
\\

]

\section{Introduction}

%
Several randomness test suites have been proposed as evaluation methods for random or pseudorandom number generators (PRNGs)\cite{NIST,TestU01}, in which randomness is tested at two levels.
The first-level test is an individual test that yields p-values as well as pass or fail results for each tested sequence, and the second-level test evaluates the results of the first-level tests.
As one of the second-level tests, the uniformity of p-values obtained by the first-level test was tested using the goodness-of-fit test. 
However, it is known that the exact distribution of p-values differs from the uniform distribution depending on the first-level test\cite{Pareschi, Haramoto, Y15}.
For the $\chi^2$ test adopted as one of the second-level tests in the test suite NIST SP80022, the effect of this difference on the test results  was analyzed, and upper limits of sample size (number of tested sequences) were proposed by F.~Pareschi et. al\cite{Pareschi}, by H.~Haramoto\cite{Haramoto}, and \cite{Y15}.
Pareschi et. al. also considered adopting the Kolmogorov-Smirnov (K-S) test as a second-level test \cite{Pareschi}, but their analysis was limited to the case where first-level tests were based on the binomial distribution.
In this study, we adopt the K-S test as the second-level test, without restricting the nature of the first-level tests. 
We analyze the effect of the deviation of the exact distribution of p-values from the uniform distribution on $[0, 1]$, which is usually assumed by the null hypothesis of randomness.
Therefore, we derive an inequality that provides an upper bound on the expected value of the K-S test statistic. 
The obtained inequality is numerically examined for a toy distribution of p-values and some of the practical first-level tests in NIST SP800-22. 
This inequality also allows us to estimate the maximal sample sizes required to pre-empt a high probability of incorrectly identifying an ideal generator as non-random.
To improve the second-level test, we propose using the K-S test based on the empirical distribution of p-values generated by the first-level test results of ideal random sequences.
In practice, we propose using pseudorandom sequences obtained from the chaotic true orbits of the Bernoulli map\cite{Saito16,Saito18}  as a substitute for such ideal random sequences.
%

\section{Second-level randomness test based on the K-S test}

%
Using the K-S test, we can test the goodness-of-fit between the empirical distribution and the reference distribution, or between two empirical distributions. 

Let $p=\{ p_i \in [0,1] | i=1,2,\cdots,m\}$ be the $m$ p-values obtained by the first-level randomness test. 
The empirical distribution with $m$ samples is defined as
\begin{equation}
\label{gpm}
G_{p,m}(x) = (1/m)\ \#\{1\leq i \leq m | p_i \leq x \},
\end{equation}
where $0\leq x \leq 1$ and $\#\{\cdot\}$ denotes the number of elements in a set $\{\cdot\}$.

Let the null hypothesis $H_0$ be $p_1,\cdots,p_m$ from the reference distribution $F$.
The reference distribution $F$ is usually assumed to be a uniform distribution $F_{unif}(x) = x \ (x \in [0,1])$.
However, there are some cases in which the exact distribution of the p-value is different from $F_{unif}$ depending on the first-level randomness test\cite{Pareschi}.
The test statistic of the one-sample K-S test with reference distribution $F$ is defined as follows.
\begin{equation}
\label{defDF}
D_F = \sqrt{m} \cdot \sup_{x\in [0,1]} |G_{p,m}(x)-F(x)| .
\end{equation}
The null hypothesis $H_0$ is accepted if
\begin{equation}
\label{oneks}
D_F \leq K(\alpha),
\end{equation}
where $K(\alpha)$ is the boundary value for the significant level $\alpha$.
This boundary value can be approximated as $K(\alpha)\simeq \sqrt{-(1/2) \log{(\alpha/2)}}$ for a large $m$ and small $\alpha$\cite{Press}.
The boundary values for $\alpha = 0.01$ and $0.0001$ are given by $K(0.01) \simeq 1.628$ and $K(0.0001) \simeq 1.949$, respectively.

\section{Inequality for the expected value of test statistic}

Let $G$ be the exact distribution for $G_{p,m}$. 
The test statistic of the K-S test with the exact reference distribution $G$ is defined as
\begin{equation}
D_G = \sqrt{m} \cdot \sup_{x\in [0,1]} |G_{p,m}(x)-G(x)| .
\end{equation}
The distribution of $D_G$ asymptotically obeys the Kolmogorov distribution under the null hypothesis if the exact reference distribution $G$ is continuous.
If the distribution of p-values of the first-level test is discrete, $G$ is not continuous but is a piecewise constant.
Following Pareschi et. al. [3], we also assume that the distribution of $D_G$ still obeys the Kolmogorov distribution, even if $G$ is piecewise constant.
In the following, we analyze the difference between the expected values of the test statistics $D_F$ and $D_G$ under this assumption. 
Applying the triangle inequality to the right-hand side of Equation \eqref{defDF}, we obtain 
\begin{align}
D_F  & =  \sqrt{m}\cdot \sup_{x\in[0,1]} |G_{p,m}(x)-G(x)+G(x)-F(x)| \nonumber \\
 & \leq  \sqrt{m}\cdot \sup_{x\in[0,1]} |G_{p,m}(x)-G(x)| \nonumber \\
 &  \ \ \ \ \ +\sqrt{m}\cdot \sup_{x\in[0,1]} |G(x)-F(x)| \nonumber \\
 & = D_G +\sqrt{m}\cdot d \ , 
\label{inequality0}
\end{align}
where 
\begin{equation}
d = \sup_{x\in[0,1]} |G(x)-F(x)| \ .
\label{def_d}
\end{equation}
This $d$ is a constant determined by the reference distribution $F$ and the exact distribution $G$ for the first-level test.

Considering the expectation with respect to the direct product of the measure determined by $G$ for inequality \eqref{inequality0}, we obtain  the inequality
\begin{equation}
E[D_F] - E[D_G] \leq  \sqrt{m}\cdot d \ .
\label{inequality1}
\end{equation}
It is known that the expected value $E[D_G]$ converges to the constant 
\begin{equation}
\mu = \sqrt{\pi / 2}\cdot \ln{2} = 0.868\cdots. 
\end{equation}
when $m\rightarrow \infty$, and the constant $\mu$ is independent of $G$ \cite{Marsaglia} .

Inequality \eqref{inequality1} implies that the difference $E[D_F] -\mu$ has an upper bound of $\sqrt{m}\cdot d$.
Note that for the $\chi^2$ test, the difference between the expected value of the test statistic based on the reference distribution that differs from the exact distribution and that based on the exact distribution is proportional to $m$\cite{Matsumoto}.

From this perspective, the K-S test is regarded as more robust to increasing sample size $m$ than the $\chi^2$ test, because the difference in the test statistics is proportional to $\sqrt{m}$ for the K-S test.
However, for the same reason, the power of the K-S test is expected to be lower than that of the $\chi^2$ test.

Furthermore, the safety of the randomness test was evaluated using inequality \eqref{inequality1}.
If the difference $\Delta$ is admissible for $E[D_F]-E[D_G]$, the maximum sample size within the difference $\Delta$ is given by $(\Delta/d)^2$.

\begin{table*}[h]
\caption{Results of the K-S test based second-level randomness tests}
\label{table1}
\begin{tabular}{r|l|ccrr|ccrr}
\hline
 &  &
   \multicolumn{4}{p{4.5cm}|}{
   (a) The one-sample K-S test with the uniform distribution
   }&
   \multicolumn{4}{p{4.5cm}}{
   (b) The two-sample K-S test with the empirical distribution
   } \\
No.  & Test name  & \multicolumn{2}{c}{p-value}& \multicolumn{2}{c|}{Pass Rate } & \multicolumn{2}{c}{p-value}& \multicolumn{2}{c}{Pass Rate  }  \\
& & mean & SD & {\tiny $\alpha\!=\!0.01 $} &  {\tiny $\alpha\!=\!0.0001$ }& mean & SD &  {\tiny $\alpha\!=\!0.01$} &  {\tiny $\alpha\!=\!0.0001$}\\
\hline
\hline
1& Frequency\,Test&0.033&0.025&8/10&10/10&0.510&0.234&10/10&10/10\\
2& Block\,Frequency\,Test &0.499&0.303&10/10&10/10&0.511&0.329&10/10&10/10\\
3& Runs\,Test&0.374&0.259&10/10&10/10&0.489&0.327&10/10&10/10\\
4&Longest\,Run\,of\,Ones\,Test&0.000&0.000&0/10&0/10&0.594&0.252&10/10&10/10\\
5& Binary\,Matrix\,Rank\,Test&0.000&0.000&0/10&0/10&0.504&0.321&10/10&10/10\\
6& Discrete\,Fourier\,Transform\,Test&0.000&0.000&0/10&0/10&0.618&0.354&10/10&10/10\\
7& Non-overlapping\,Template\,Matching\,Test\,(1)&0.394&0.314&10/10&10/10&0.656&0.303&10/10&10/10\\
8& Overlapping\,Template\,Matching\,Test&0.000&0.000&0/10&0/10&0.636&0.260&10/10&10/10\\
9& Maurer's\,"Universal\,Statistical"\,Test&0.000&0.000&0/10&0/10&0.439&0.240&10/10&10/10\\
10& Linear\,Complexity\,Test&0.064&0.121&6/10&10/10&0.489&0.291&10/10&10/10\\
11& Serial\,Test\,(1)&0.415&0.131&10/10&10/10&0.520&0.170&10/10&10/10\\
12& Approximate\,Entropy\,Test&0.000&0.000&0/10&0/10&0.394&0.347&9/10&10/10\\
13& Cumulative Sums Test (1)&0.089&0.138&7/10&10/10&0.409&0.247&10/10&10/10\\
\hline

\end{tabular}
\end{table*}

\section{Two-sample K-S test with ideal empirical distribution}

A simple method to improve the K-S test based second-level test involves the use of the statistic $D_G$ instead of $D_F$ if the exact distribution $G$ is known for the target first-level test. 
In this case, we can obtain test statistics without the error effect. 
However, it is not always possible to compute the exact distribution for a given first-level test.
Therefore, as another method, we examine a method that uses the empirical distribution of p-values obtained from the first-level test for ideal random sequences as the reference distribution. 

Let  $q = \{q_i \in [0,1] | i = 1,2,\cdots,m'\} $ be the $m'$ p-values  obtained by the first-level test for ideal or nearly ideal random sequences.
By the definition, the distribution of $q$ obeys $G$.  
Similar to Equation \eqref{gpm}, the empirical distribution of $q$ is defined as 
\begin{equation}
G_{q,m'}(x) = (1/m')\ \#\{1\leq i \leq m' | q_i \leq x \}.
\end{equation}

By using the two-sample K-S test, the goodness-of-fit between the empirical distribution $G_{p,m}$ and $G_{q,m'}$ is also tested as a second-level randomness test.
The test statistic of this two-sample K-S test is defined as
\begin{equation}
D_{G_{q,m'}} = \sqrt{\frac{m\cdot m'}{m+m'}} \cdot \sup_{x\in [0,1]} |G_{p,m}(x)-G_{q,m'}(x)| .
\end{equation}
For the two-sample K-S test, the null hypothesis 
that $p_1,\cdots,p_m$, and $q_1,\cdots,q_m'$ are from the same exact distribution $G$ is accepted if
\begin{equation}
\label{twoks}
D_{G_{q,m'}}  \leq K(\alpha)
\end{equation}
for the significance level $\alpha$.

In this study, we propose to construct an empirical distribution $G_{q,m'} $ using the chaotic true orbit of the Bernoulli map\cite{Saito16, Saito18}. 
The dynamical system given by the Bernoulli map is defined as 
\begin{equation}
x_{i+1} = 2 x_i \mod 1, 
\end{equation}
where $x_i \in [0,1)$ and $i=0,1\cdots$.
By providing an irrational algebraic number as an initial state $x_0$, we can generate a chaotic true orbit $x_i$ with infinite precision. 
Then, we can obtain the binary sequence $\varepsilon = \varepsilon_0, \varepsilon_1, \cdots$ by assigning  
\begin{equation}
\varepsilon_i = \left\{
\begin{array}{cc}
0 & (x_i < 1/2) \\
1 & (x_i \geq 1/2)
\end{array} \right.  .
\end{equation}
This binary sequence $\varepsilon$ corresponds to the binary expansion of the initial state $x_0$. 
See \cite{Saito16} and \cite{Saito18} for mathematical support for the good statistical qualities of $\varepsilon$.

\section{Numerical results}

\subsection{Examples of second-level tests  based on the K-S test}

As a first numerical experiment, two second-level tests based on the K-S test were applied to some of the first-level tests in NIST SP800-22. 
One second-level test was based on the one-sample K-S test with the reference distribution $F_{unif}$, and the second is the second-level test based on the two-sample K-S test with the empirical distribution that was separately prepared.
We performed these second-level tests ten times, wherein, for each second-level test, we used the p-values obtained by applying the first-level test to $m = 10^6$ sequences with length $n = 10^6$.
The tested sequences were  generated by the Mersenne twister-based PRNG.
The empirical distribution $G_{q,m'}$ used as a reference was constructed based on the results of  the first-level tests for the PRNG based on the chaotic true orbit of the Bernoulli map with $m'=10^7$ and $n=10^6$.

The results of the one-sample K-S test and the two-sample K-S test are shown in columns (a) and (b) of Table 1, respectively.
Here, the mean and the standard deviation of the ten obtained p-values, and the pass rate of the number of passes divided by ten are shown for each randomness test.
For the first-level tests of Nos. 7, 11, and 13,  which consist of several tests, the result for one test is only shown as an example.  
The random excursions test and the random excursions variant test were excluded because the number of obtained p-values varied depending on the tested sequences.
The results of the one-sample K-S test with a uniform distribution completely failed for the first-level tests of Nos. 4, 5, 6, 8, 9, and 12.
However, almost all the results of the two-sample K-S test with the empirical distribution were successful. 
These results suggest an improvement in the second-level test using the two-sample K-S test with the empirical distribution constructed using high-quality PRNG.

\subsection{Examination of the derived inequality}

\begin{figure}[t]
\centering
\includegraphics[scale=0.25]{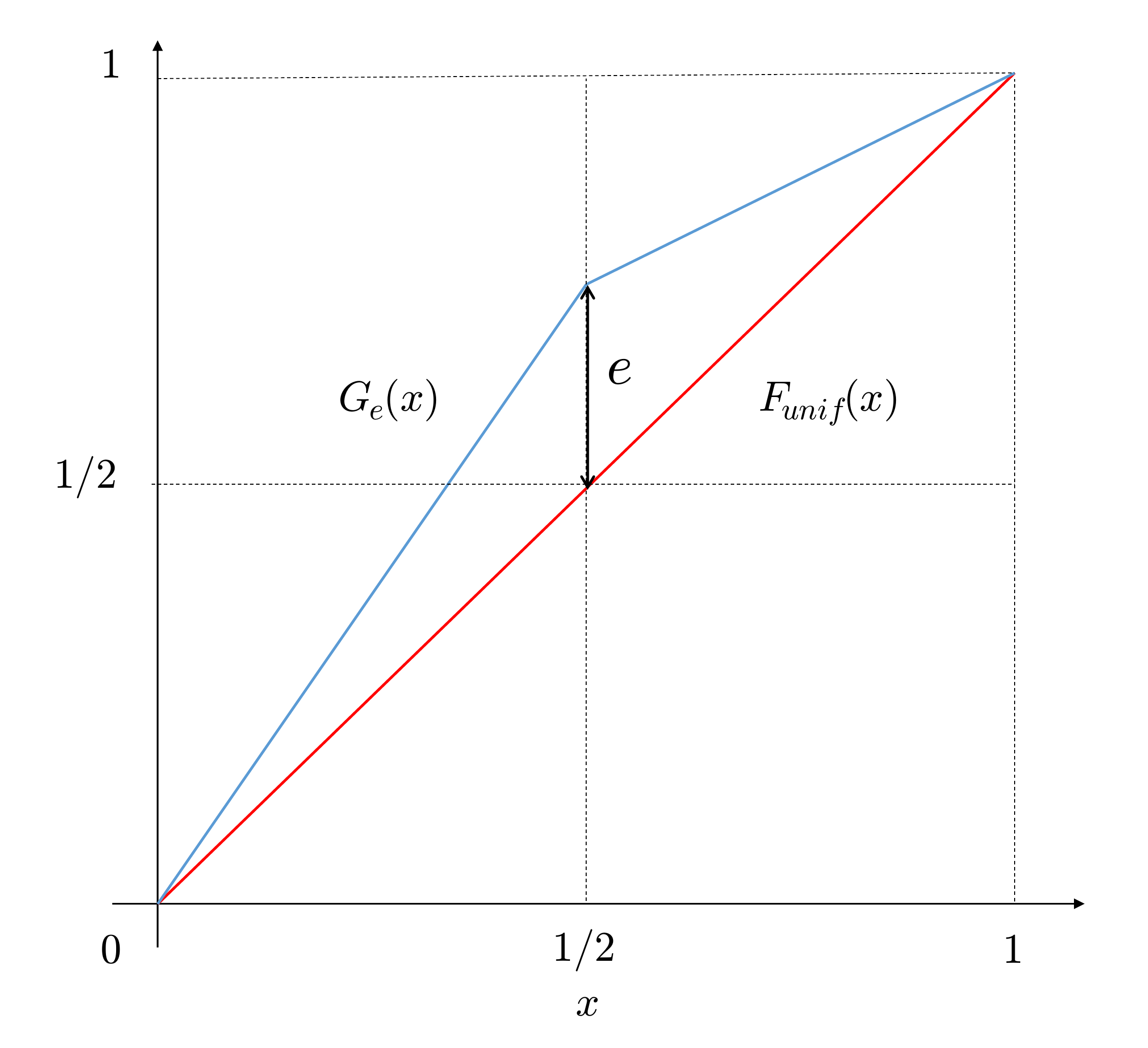}
\caption{ Examined distribution of $G_e$ and $F_{unif}$.}
\label{Fig1}
\end{figure}

To examine the inequality \eqref{inequality1}, we numerically analyze the difference between test statistics $D_F$ and $D_G$ for a particular distribution $G$ under the reference distribution $F=F_{unif}$.
As a toy model, we consider the exact distribution $G_e$, which is a piecewise linear function, given by
\begin{equation}
G_e(x) = \left\{
\begin{matrix}
(1+2e)x      & x\in [0,1/2] \\
(1-2e)x+2e & x\in (1/2,1] 
\end{matrix}
\right. \ ,
\end{equation}
where $|e|<1$ . 
The graph of $G_e$ is shown in Fig. \ref{Fig1}. 
The constant $d$ in Equation \eqref{def_d} for $G_e$ and $F_{unif}$ is equal to $e$.

\begin{figure}[t]
\centering
\includegraphics[scale=0.3]{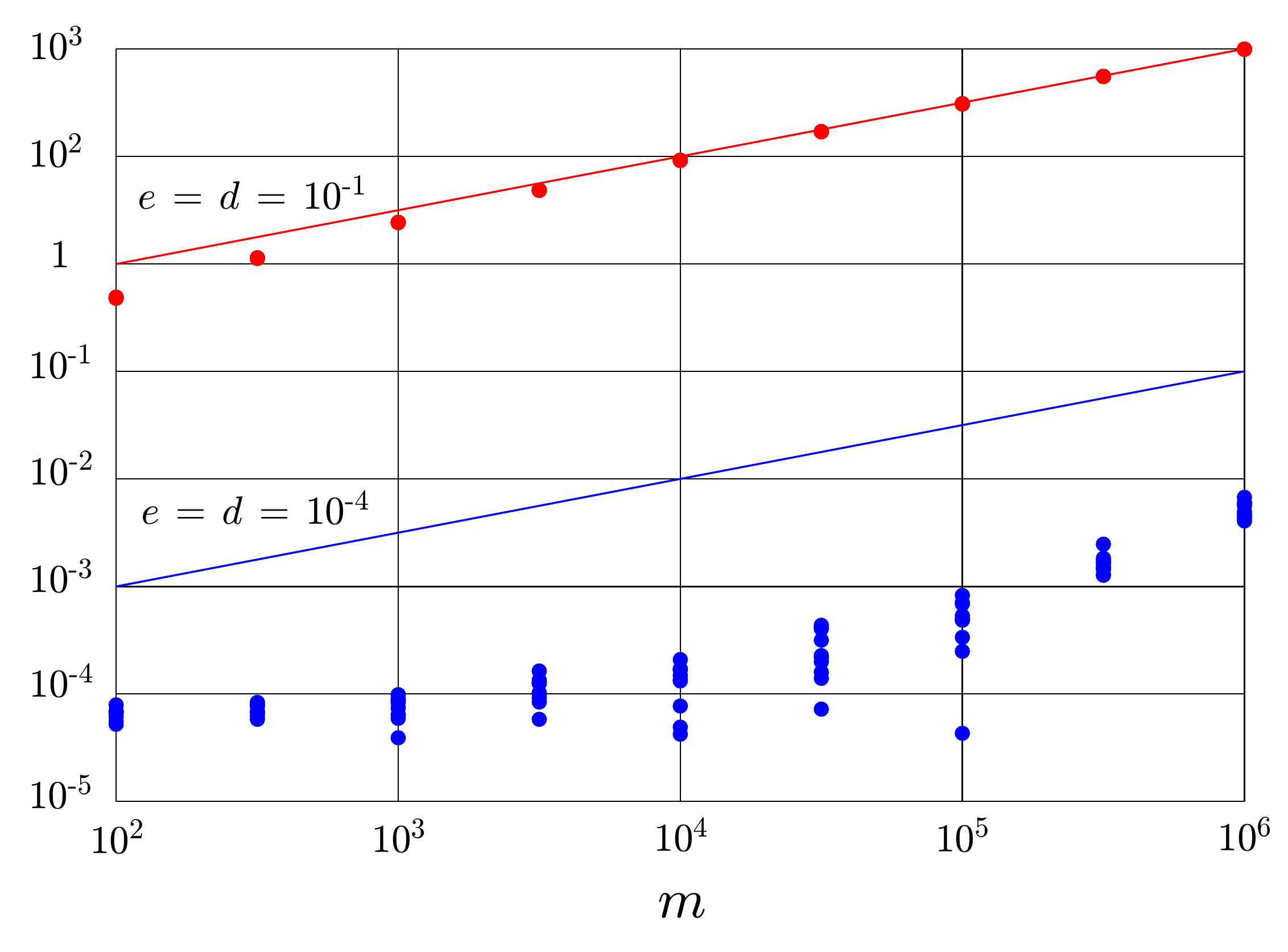}
\caption{ Differene $\Delta_m = \overline{D_{G}}-\overline{D_{F}}$ for $G_e$ and $F_{unif}$.}
\label{Fig2}
\end{figure}

For a given sample size $m$ and constant parameter $d$, we randomly generate $p_1, p_2, \cdots, p_m \in [0,1]$ that obeys the distribution $G_e$ and calculate $D_{F}$ and $D_{G}$ for $10^4$ times. 
Then, we obtain the mean values $\overline{D_F}$ and $\overline{D_G},$ and $\Delta_m = \overline{D_{F}}-\overline{D_{G}}$, respectively.
In Fig. \ref{Fig2},  $\Delta_m$ (circles) and $\sqrt{m}\cdot d$ (solid line) are shown for the cases $e=d=10^{-1}$ and $10^{-4}$. 
Here, ten samples of $\Delta_m$ are plotted for each $m$.
As a result, $\Delta_m$ is less than $\sqrt{m}\cdot d$ for both cases and converges to $\sqrt{m}\cdot d$ with increasing $m$ for $e=d=10^{-1}$.
This result is consistent with the inequality \eqref{inequality1}.

\subsection{Safe sample sizes for the frequency test and the binary matrix rank test}

Here, we analyze the frequency test and the binary matrix rank test shown in Table 1 as examples.
The frequency test was analyzed by Pareschi et. al. as an example of tests based on binomial distribution.
As a different example, we analyzed the binary matrix rank test based on the trinomial distribution.
The binary matrix rank test also failed for the one-sample K-S test with a uniform distribution.
For these two tests, we calculated the exact distributions for the sequence length $n=10^6$ and obtained the exact value of the constant $d$\cite{Y18}.
The statistics $\overline{D_{F}}$ and $\overline{D_G}$, and their difference $\Delta_m$ were also calculated from the test results shown in column (a) of Table 1.
Results are shown in Table 2.
The range of the standard error of the mean (SEM) is also shown. 
The difference $\Delta_m$ is less than $\sqrt{m}\cdot d$ for both tests, and these results are consistent with the inequality \eqref{inequality1}.

For the safety of these tests, we can obtain the maximum sample size for the given admissible difference $\Delta$ of the expected values of $D_F$ and $D_G$, as mentioned in Section 3. 
For example, if $\Delta=0.1628$, which is 10\% of the boundary value $K(0.01)$, is admissible, the maximum sample size is $15,703$ for the frequency test and $1,693$ for the binary matrix rank test. 
Furthermore, the sample size $m=10^3$, which is the recommended parameter of NIST SP800-22, is safe if $\sqrt{m}\cdot d \simeq 0.025$ is admissible for the frequency test, and $\sqrt{m}\cdot d \simeq 0.153$ is admissible for the binary matrix rank test.

\begin{table}[t]
\caption{The difference between the mean values $\overline{D_F}$ and $\overline{D_G}$ for the frequency test and the binary matrix rank test}
\renewcommand{\arraystretch}{1.5}
\begin{tabular}{ccc}
\hline
   & Frequency\,Test  & Binary\,Matrix\,Rank\,Test  \\
\hline
\hline
$\overline{D_F}$  &  1.319$\pm$0.114  &  5.082$\pm$0.107 \\
$\overline{D_G}$  &  0.863$\pm$0.057  &  0.840$\pm$0.093 \\
$\Delta_m\!=\!\overline{D_F}\!-\!\overline{D_G}$ & 0.456$\pm$0.100 & 4.242$\pm$0.120 \\
\hline
\hline
$\sqrt{m}\cdot d$ & 0.798 & 4.860 \\
\hline
\end{tabular}
\renewcommand{\arraystretch}{1}
\end{table}

\section{Conclusion}

In this work we derived an inequality that provides the upper bound on the difference of the expected values of the test statistics for the K-S test based second-level randomness test.
The derived inequality was numerically examined and consistent results were obtained.
In addition, we examined the second-level test that uses the two-sample K-S test with the nearly ideal empirical distribution constructed from the PRNG based on the chaotic true orbit for several randomness tests in NIST SP800-22.
These results are expected to prove useful for evaluating the safety of the randomness test using the K-S test. 
We intend to perform an analysis of the other goodness-of-fit tests, such as the Cr\'{a}mer-von-Mises test and the Anderson-Darling test, in future work.

\section*{Acknowledgement}
This work was supported by JSPS KAKENHI Grant Numbers 16KK0005, 17K00355. The computation was carried out using the computer resources offered under the category of General Projects by the Research Institute for Information Technology, Kyushu University.

\end{document}